\documentclass[12pt]{article}
\usepackage[letterpaper,margin=1in]{geometry}
\usepackage{amsmath}
\usepackage{amssymb}
\usepackage{amsthm}
\usepackage{dsfont}
\usepackage{natbib}
\usepackage{graphicx}
\usepackage{color}
\usepackage{setspace}
\usepackage{subcaption}
\usepackage{tabularx}
\usepackage{pdflscape}

\author{Jacob Mortensen, Luke Bornn}

\begin{document}
\title{Estimating locomotor demands during team play from broadcast-derived tracking data}
\maketitle

\begin{abstract}
The introduction of optical tracking data across sports has given rise to the ability to dissect athletic performance at a level unfathomable a decade ago. One specific area that has seen substantial benefit is sports science, as high resolution coordinate data permits sports scientists to have to-the-second estimates of external load metrics, such as acceleration load and high speed running distance, traditionally used to understand the physical toll a game takes on an athlete. Unfortunately, collecting this data requires installation of expensive hardware and paying costly licensing fees to data providers, restricting its availability. Algorithms have been developed that allow a traditional broadcast feed to be converted to x-y coordinate data, making tracking data easier to acquire, but coordinates are available for an athlete only when that player is within the camera frame. Obviously, this leads to inaccuracies in player load estimates, limiting the usefulness of this data for sports scientists. In this research, we develop models that predict offscreen load metrics and demonstrate the viability of broadcast-derived tracking data for understanding external load in soccer.

\end{abstract}

\section{Introduction}

In order to reduce fatigue, prevent injury, and improve performance, sports scientists seek to monitor the physical impact that participation in training and competition has on an athlete. A number of different metrics, broadly referred to as load metrics, have been used to try and quantify the intensity of a given activity for an athlete. These metrics consist of two general categories: internal and external. \cite{Halson2014} defines internal load as ``the relative physiological and psychological stress imposed'' on an athlete. Internal load measures are not treated in this work and we make no further note of them other than to mention that they exist and include, for example, individual reporting of perceived exertion, heart-rate-derived training impulse, and summated-heart-rate-zones \citep{Borresen2008,McLaren2018}. External load is defined as ``the work completed by the athlete, measured independently of his or her internal characteristics'' \citep{Halson2014}. Metrics in this category include distance measures (both total distance over a training session or match and distance traveled stratified by intensity of the activity) \citep{Coutts2010,Rampinini2009,Dalen2016,McLaren2018} and acceleration-derived measures \citep{Dalen2016,Delaney2016,McLaren2018,Nicolella2018,Boyd2011}. Existing methods for accurately capturing external load metrics consist of an athlete wearing a device, whether that be a global positioning system (GPS) \citep{Sykes2013,Mullen2019} or local positioning system (LPS) tracker \citep{Vazquez-Guerrero2019}, or an accelerometer \citep{Boyd2011}, which can give accurate readings of instantaneous velocity and acceleration. However, it is not always possible for an athlete to wear such a device, and so optical tracking data has been used as an alternative means to capture measures of external load \citep{Gregson2010}. 

Since its initial introduction in soccer by Prozone in 1999 \citep{Medeiros2017}, multicamera optical tracking data has spread across a variety of sports, providing detailed location data for players multiple times per second. The prevalance of this data allows sports analysts to move beyond the limitations of box score statistics, capturing tactics, strategy, and nuance in a manner that is impossible with the simple enumeration of discrete events. This type of data allows analysts to move from simply noting when an assist occurred to answering questions like:  ``where were the assisting players teammates when the assist occurred?'', ``what type of defensive pressure was the assist made under?'', and ``what types of actions increase the likelihood of an assisted goal?'' The value of this data is shown clearly by \cite{Cervone2016}, who introduced the idea of expected possession value (EPV) by using tracking data from the National Basketball Association to show how the value of a possession evolves over time, accounting for contextual information like which players were on the court, where they were located, and potential actions. This EPV framework was later extended to soccer by \cite{Fernandez2019}, who were able to use it to show, as just one example, where space was being created on the pitch and how space creation increased or decreased the value of a possession. 

Tracking data is useful for sports scientists because it allows them to derive metrics related to distance, speed and acceleration that serve as a proxy for the stress placed on a player's body as a result of their athletic performance. Because the equipment to produce this data is fixed in the arena and not hampered by either league rules or player cooperativeness, this results in a much more expansive sample of player load than is possible via wearable devices. However, despite the distinct advantages provided by multicamera optical tracking data two extant issues remain: exclusivity and sparsity. This data is exclusive because obtaining it requires installation of expensive hardware and paying large licensing fees, restricting its availability to only the most elite leagues. The data is sparse in the sense that it has become widespread only in recent years, preventing historical comparison. As a result, any ability to draw conclusions about load metrics and their relationship to health outcomes is limited. Broadcast-derived tracking data has the capacity to overcome both of these problems because it allows coordinates to be extracted from regular broadcast video using computer vision techniques \citep{Lu2013}. This eliminates the need to install special cameras and has the potential to provide x-y coordinate data for any game with a video feed. 

Despite its exciting possibility, broadcast-derived tracking data comes with one glaring issue: coordinate data is only available for a player as long as they are within the camera frame. Our purpose in this paper is to assess the viability of broadcast-derived tracking data for estimation of a variety of external load metrics commonly used across sports. Specifically, our focus is on estimating load metrics during the time that a given player is offscreen. We do this by using games for which complete multicamera tracking data is available, and manually censoring observations to emulate the broadcast-derived tracking data. Approaching the problem in this way allows us to establish a ground truth, answering definitively, given that the broadcast tracks are accurate, whether or not broadcast-derived tracking data can be used to assess external load. 

\section{Methods}\label{sec:methods}

\subsection{Data}\label{sec:data}

Our data comes from 18 of the 19 home games played by Chelsea FC in the 2014-15 English Premier League and includes information for 248 players (this number does not include goalkeepers, which are excluded from our analysis). %, with the 13 September 2014 match against Swansea City being the only exception. 
The data contains complete x-y coordinate data for all players in each game at a frequency of 10 measurements per second, providing a continuous track for a player for the entire time he was in the game, with a break at match halftime. Additionally, the data includes event information, which consists of the location and description of actions such as a touch, pass or tackle. In order to emulate the broadcast derived tracking data while retaining true offscreen values, we simulate a camera track by linearly interpolating between event locations and place a $40 \times 40$ meter window centered on the camera track. Player locations outside of the window are treated as unobserved, and metrics calculated from these censored tracks are the values we predict in this paper. 
Each time the track transverses the edge of the camera window, we split the multicamera track into a new segment, which we refer to as a subtrack, and assign it a unique ID. This process results in 149,680 subtracks generated from an original 820 tracks, giving an average of 8315.6 (+/- 411.9) subtracks per game from a median of 45.5 (range=44-47) original player tracks per game. The median subtrack length is 17.0 m (range=0.0-830.1 m) with a median time of 8.1 s (range=0.1-556.7 s). Subtrack information is augmented by player position information scraped from transfermarkt.com, with each player classified as either a defender (n = 85), midfielder (n = 85), or forward (n = 78). 

In order to make predictions a variety of features were constructed from the tracks, at both the subtrack level and aggregated to the game level. For each subtrack, we record the x and y location for where the player left and re-entered the camera window and calculate the Euclidean distance and time elapsed between them. Distance, velocity, and acceleration are calculated for each 0.1 second interval, and used to calculate the load metrics detailed below as well as average velocity and average absolute acceleration in the 2 second intervals preceding and following each subtrack. The raw accelerations exhibit some unrealistic values, with some instantaneous accelerations greater than 50 $\text{m}/\text{s}^{-2}$, so in order to reduce this noise, we smooth the accelerations using a Nadaraya-Watson kernel smoother \citep{Nadaraya1964,Watson1964}. At the game level, most of the features consist of the load metrics calculated for the observed portion of the game, but we also calculate the total time in seconds that a player was censored, percent of playing time a given player was censored, and their average observed velocity. 

\subsection{Player load metrics}
\label{subsec:player_load_metrics}

The suite of external load metrics we consider in this work were selected because of their use throughout the literature (see, for example, \cite{Varley2013,Gabbett2012,Dwyer2012,Johnston2014a,Dalen2016,Borresen2008,McLaren2018}). Because specific definitions of load metrics vary wildly throughout the literature, we provide definitions for the metrics we use in Table~\ref{tab:player_load_metrics}. Note that although we are only making predictions for the eight load metrics outlined in Table~\ref{tab:player_load_metrics} that they fall into the three broad categories of distance, velocity, and acceleration derived measures. In general, any external load metric that falls into one of these categories can be calculated from broadcast-derived tracking data, though prediction accuracy for specific censored metrics should be assessed individually.

\begin{figure}
  \includegraphics[width=\textwidth]{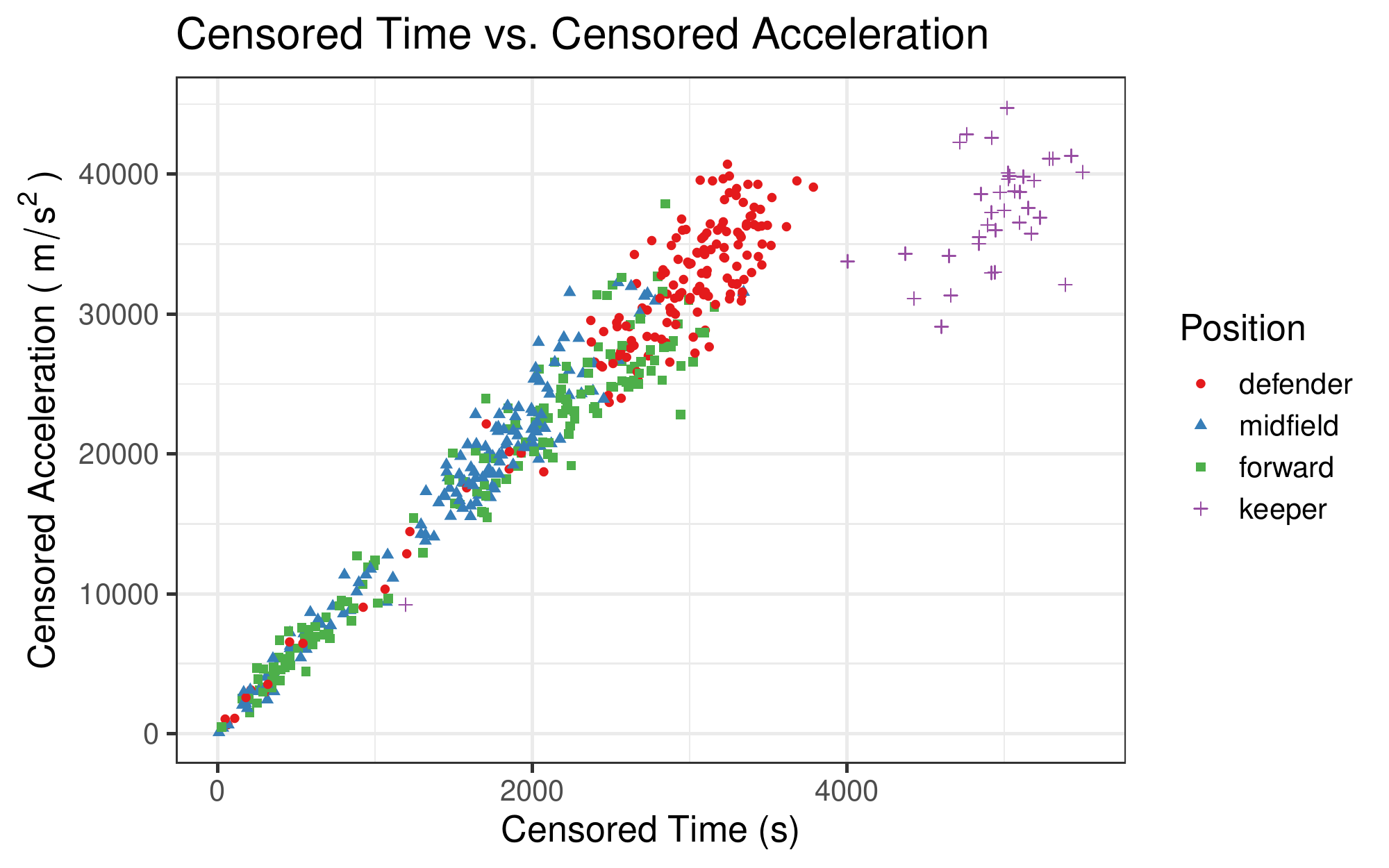}
  \caption{Censored time versus censored total acceleration. Note that once keepers are removed from the data, the relationship between censored time and censored total acceleration is almost exactly linear.}
  \label{fig:time_v_acc}
\end{figure}

\begin{figure}
  \includegraphics[width=1\textwidth]{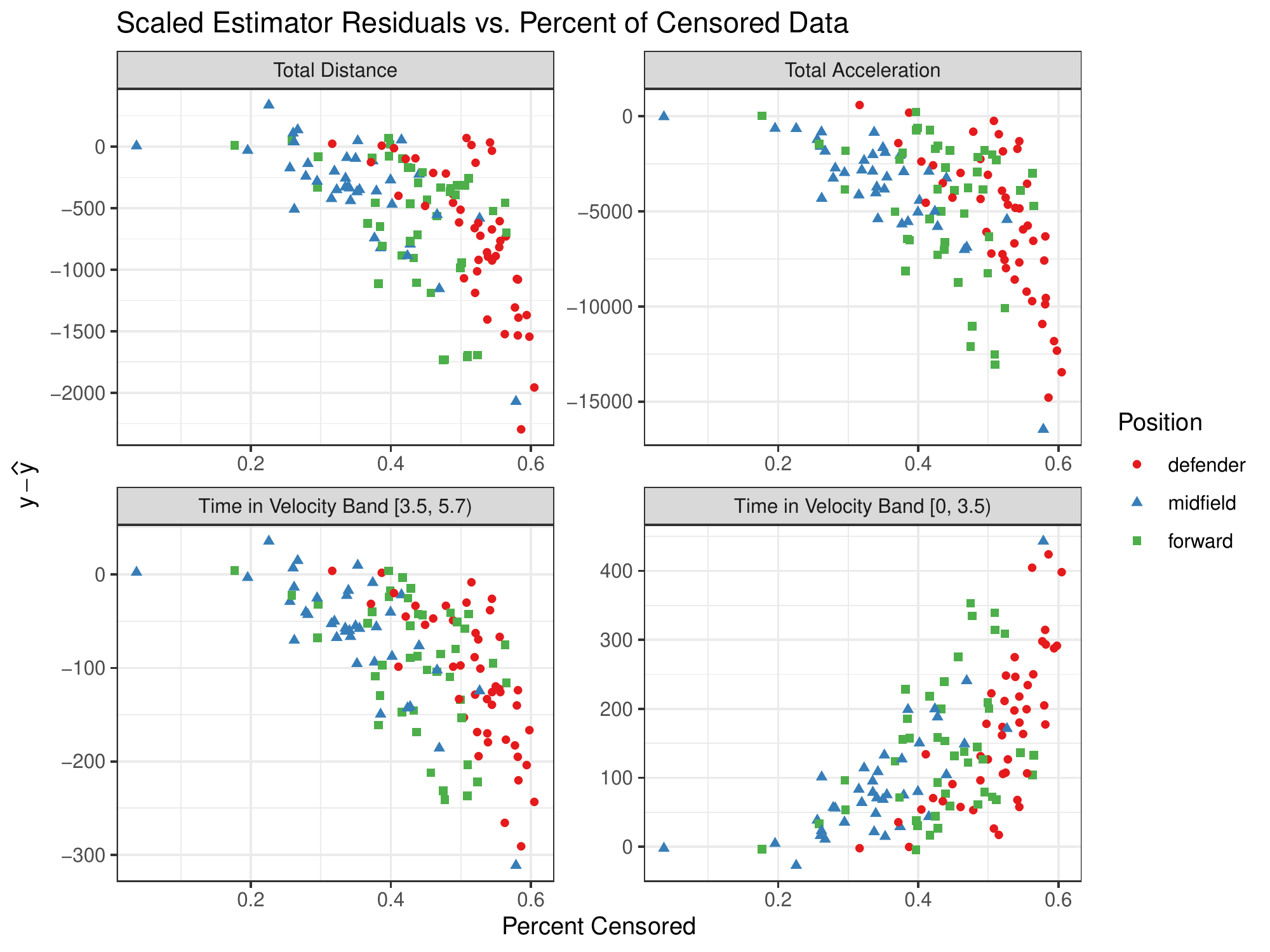}
  \caption{Residuals for the scaling estimator in Section \ref{subsec:player_load_metrics} versus the percentage of data that is censored. The scaling estimator consistently underestimates the amount of time spent in the slowest velocity band, whereas it consistently overestimates values for the other load metrics, resulting in negative residuals. This is clear evidence that the way players move on camera differs systematically from how they move offscreen.}
  \label{fig:scaling_estimator}
\end{figure}

Exploratory analysis of this data and the calculated metrics reveals two patterns worth highlighting. The first is that there is a very strong correlation between the amount of censored playing time in a game and most of the censored load metrics, as shown in Figure \ref{fig:time_v_acc} for total acceleration. When keepers are removed from the data, the values for Pearson's correlation coefficient between elapsed time for a censored subtrack and most of the other metrics range between 0.596 and 0.998, the exceptions being peak velocity and acceleration density. This suggests that in many cases fairly good estimates can be obtained by simply regressing the metric on censored subtrack time. The second pattern becomes clear if we assume that the censored data is missing completely at random (MCAR) \citep{Rubin1974}; that is, we assume that there is no relationship between the pattern of missingness and the values of the observed and censored load metrics. To illustrate, consider total distance (though the following relationship holds for the other player metrics). Under this assumption, the ratio of observed distance, $D_o$, to censored distance, $D_c$ is equivalent to the ratio between observed time, $T_o$, and censored time $T_c$, or in mathematical notation, $$\frac{D_c}{D_o} = \frac{T_c}{T_o}.$$ This in turn implies that we can estimate $D_c$ by setting $D_c = D_o \frac{T_c}{T_o}$. Because we are simply scaling the observed metric value by the ratio of censored to observed time, we refer to this as a scaling estimator. Despite its appealing simplicity, examination of the residuals, shown for four of the metrics in Figure \ref{fig:scaling_estimator}, reveals that the assumption that data is MCAR is incorrect. We see that residuals become increasingly negative (censored values are overestimated) as the amount of censoring increases for most of the metrics, with the exception being the amount of time spent in the slowest velocity band. The systematic differences between player movement on- and off-camera demonstrated in these plots can be summarized simply as ``players move faster when on camera.'' 

\begin{table}[!htbp] \centering 
  \caption{External load metric definitions} 
  \label{tab:player_load_metrics} 
  \begin{tabularx}{\textwidth}{l l X}
  \\[-1.8ex]\hline 
  \hline \\[-1.8ex] 
    Category & Metric & Definition \\
    \hline \\[-1.8ex] 
    Distance & Total distance & Sum of the distance travelled by an athlete. \\
    & High speed distance &  Sum of the distance travelled by an athlete with speed between 3.5 and 5.7 m/s. \\
    & Very high speed distance &  Sum of the distance travelled by an athlete with speed greater than 5.7 m/s. \\
    Velocity & Time spent in velocity band $[x,y)$ &  Number of seconds spent with velocity $(m/s)$, $v$, in the interval $x \le v < y$. Intervals considered are $[0,3.5)$, $[3.5,5.7)$, and $[5.7, \infty)$, based on the work of \cite{Dwyer2012}. \\
    & Peak $x$-second velocity & Max velocity of average velocities calculated over $x = 1, 3, 5,$ and $10$ second rolling windows. \\
    Acceleration & Total Acceleration & Sum of the absolute values of acceleration at 0.1 second intervals. \\
    & Acceleration density & Mean acceleration. \\
    & Time spent in acceleration band $[x,y)$ & Number of seconds spent with acceleration $(m/s^2)$, $a$, in the interval $x \le a < y$. Intervals considered are $[0.65, 1.46)$, $[1.46, 2.77)$, and $[2.77, \infty)$, based on the work of \cite{Johnston2014a} \\

  \hline \\[-1.8ex] 
  \end{tabularx}
\end{table}

\subsection{Models}\label{sec:models}

% I want to show:
% - we used xgboost, both random forest and linear 
% - we compare models at the subtrack and game level
% - what variables do we consider?
% - fit models on first 13 games, use last 5 as test set
% - how do we compare models? (rmspe, out of sample R2)

In order to establish a baseline level of comparison, we use a linear regression model for each load metric of the form $$y_i = \beta_0 + \beta_1 T_i + \beta_1 x_i + \epsilon,$$ where $i$ indexes the player-match combination, $y_i$ is the value of the censored metric at the game level, $T_i$ is the amount of censored time for the game, $x_i$ is the value of the observed metric at the game level, and $\epsilon$ is a normally distributed error term. The remaining models that we compare are all fit using gradient boosting \citep{Friedman2001} as implemented by the \texttt{xgboost} package \citep{Chen2016} within the statistical programming language \texttt{R} \citep{RCoreTeam2019}. Gradient boosting is an ideal tool for this application for several reasons: it fits a model, then iterates, fitting a model on the residuals of each previous model until no further improvements can be made, then averages all of the models together, yielding very accurate predictions; it is flexible because it allows the use of both linear and nonlinear (tree-based) boosters; and it has the convenient feature of performing automatic variable selection, permitting us to take a ``kitchen-sink'' approach and include all potential predictors for consideration in the model. Additionally, as implemented in \texttt{xgboost} it is extremely fast, allowing models to be fit to a large amount of data in just a short span of time.

We compare two different approaches to obtaining game level estimates: predicting the values for each subtrack individually and then aggregating them to the game level versus predicting game level metrics directly. We combine these two levels of estimation with three different models: a linear model with no interactions, a linear model that considers all two-way interactions, and a random forest model. We train each model on the first thirteen games of the season, reserving the final five games for testing, which results in 361 player-match observations in the training set and 138 player-match observations in the test set. We assess model performance by comparing root mean square predictive error (RMSPE), defined $\text{RMSPE} = \sqrt{\sum (y_i - \hat{y}_i)^2 / n}$, where $\hat{y}_i$ is the predicted value for observation $y_i$ and $n$ is the number of observations, and the coefficient of variation (CV), defined $\text{CV} = RMSPE /  \bar{y}$, where $\bar{y} = \sum_{i=1}^n y_i / n$. In this context CV is useful because it tells us how large the errors are relative to the values themselves, giving an indication of how much the overall variance in the data was reduced by the given model. 

The predictors included in the various models are denoted in Table~\ref{tab:model_predictors}. All numeric variables were centered and scaled so that they have a mean of 0 and standard deviation of 1, which, for the linear models, allows us to determine relative significance of each predictor in the model simply by comparing the size of their associated coefficients. 

Here we briefly note that 100\% of the peak $x$-second velocity values in the 18 games in our data set occur within the camera window, and as such, there is no need to try and estimate these values for the censored tracks. While it may not always be the case that peak velocities are always observed, this suggests that any exceptions will be rare, and so peak velocity is omitted from the subsequent analysis.

\begin{table}[!htbp] \centering 
  \caption{Model predictors. Inclusion for consideration in subtrack or game level models is indicated by the x. All variables that begin with ``observed'' are measured at the game level, so an x in the subtrack model column for ``observed total acceleration'' means that the sum of the absolute value of accelerations for the entire game is used as a predictor when estimating individual subtrack outcomes.} 
  \label{tab:model_predictors}
  \footnotesize
  \begin{tabular}{l c c}
  \\[-1.8ex]\hline 
  \hline \\[-1.8ex] 
    Predictor & Included at subtrack level & Included at game level \\
    \hline \\[-1.8ex] 
    player position & x & x \\
    offscreen time & x &  \\
    censored total time &  & x \\
    offscreen distance & x &  \\
    observed total distance &  & x \\
    average velocity in previous two seconds & x & \\
    average velocity in following two seconds & x & \\
    average absolute acceleration in previous two seconds & x & \\
    average absolute acceleration in following two seconds & x & \\
    observed average acceleration & x & x \\
    observed total acceleration & x & x \\
    observed average velocity & x & x \\
    observed high speed distance & x & x \\
    observed very high speed distance & x & x \\
    observed time in velocity band $[0, 3.5)$ & x & x \\
    observed time in velocity band $[3.5, 5.7)$ & x & x \\
    observed time in velocity band $[5.7, \infty)$ & x & x \\
    observed time in acceleration band $[0.65, 1.46)$ & x & x \\
    observed time in acceleration band $[1.46, 2.77)$ & x & x \\
    observed time in acceleration band $[2.77, \infty)$ & x & x \\
  \hline \\[-1.8ex] 
  \end{tabular}
\end{table}

\section{Results}\label{sec:results}

% What do I want to show here?
% - subtrack estimation is more accurate
% - in general, R^2 values indicate we are getting very good estimates relative to the overall variance
% - display at least one residual plot and show the impact of percent censored on the response estimates
% - discuss which variables are most important

The full results for each of the predicted metrics at the subtrack and game levels are shown in Tables~\ref{tab:subtrack_results} and \ref{tab:game_results}, respectively. In all cases, estimating the player metrics at the subtrack level and then aggregating to obtain game level estimates outperforms making predictions purely at the game level, as seen in the lower RMSPE and CV values. For seven of the eleven player load metrics under consideration, the linear model with interactions performs the best, though the random forest has the lowest RMSPE and CV in three cases. Predicting acceleration density is the only case where the linear model with no interactions results in the best predictions. 

CV values are less than or equal to 0.10 for six of the eleven models, indicating a significant reduction in standard error relative to the overall size of the response. The largest CV is 0.31, for both very high speed distance and time in velocity band $[5.7, \infty)$, both outcomes for which the nonlinear model performs the best. This result can be explained when we consider that the correlations of these two metrics with censored total time are 0.599 and 0.605, respectively, indicating that these predictions do not benefit from the very strong relationship with censored total time that the other metrics do. Considering the RMSPE values themselves helps us understand just how well each metric is being predicted. For example, despite its CV of 0.31, the RMSPE for time in velocity band $[5.7, \infty)$ is still only 6.4 seconds. The RMSPE for total distance is 183 meters, minimal error considering players on average travel 3524 meters in each game.

\begin{figure}
  \includegraphics[width=\textwidth]{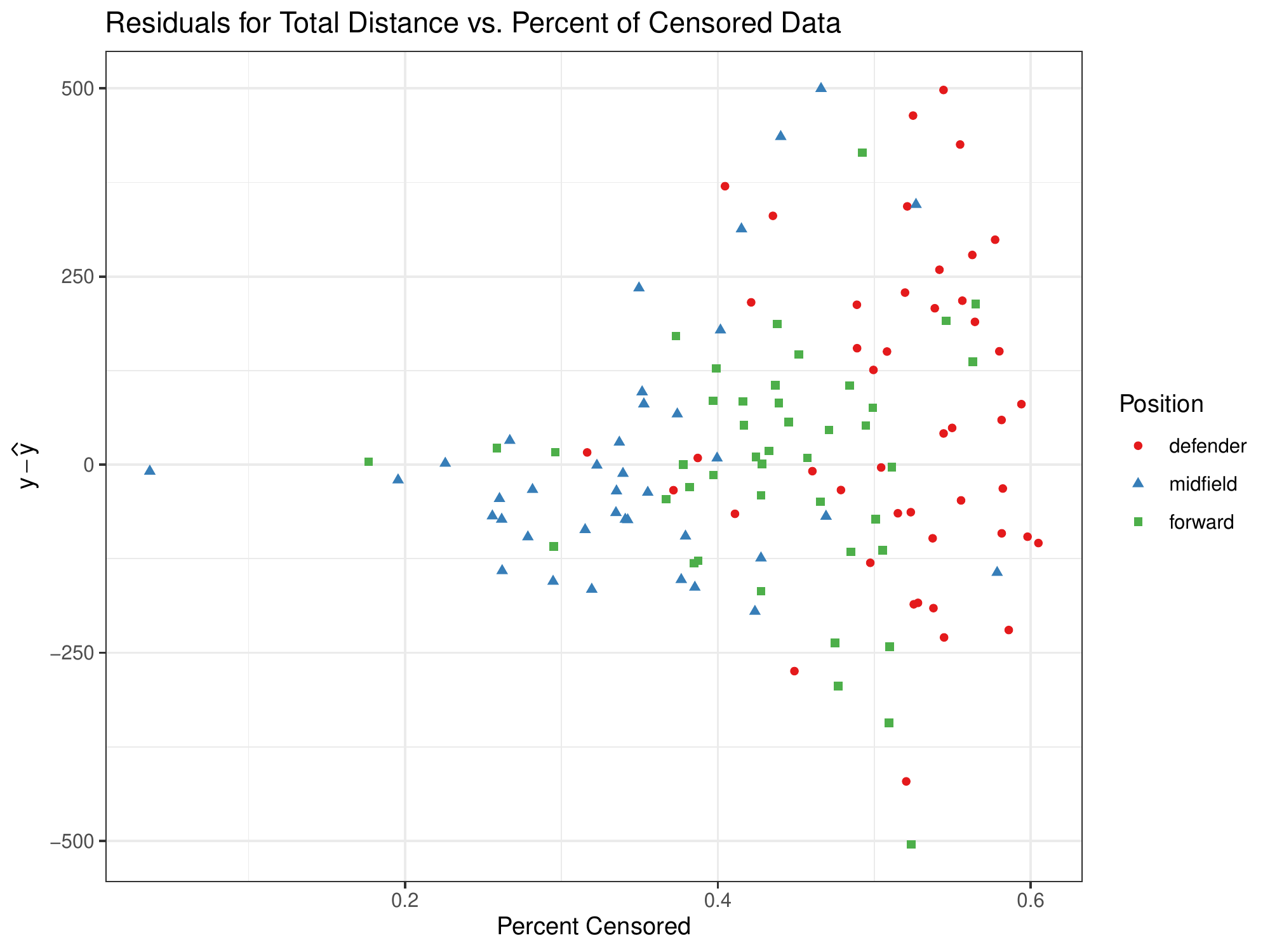}
  \caption{Residuals for total distance predictions versus percent of total data that is censored.}
  \label{fig:residuals}
\end{figure}

An examination of the residuals for a given response variable against the percent of data that is censored, shown for total distance in Figure \ref{fig:residuals}, is illuminating. Unsurprisingly, we see that the variability in the predictions increases with the amount of censored data, but in general the predictions appear unbiased, and even when as much as 50\% of the data is missing, the range of the residuals being approximately (-500, 500) indicates a significant reduction in variability when compared to the empirical standard deviation for the response of 1506 meters. Figure \ref{fig:residuals} also shows a clear demarcation in the amount of censoring across the various positions, with defenders experiencing the most time off camera and midfielders spending the most time within the camera frame. 

Due to the nature of gradient boosting, our ability to make inference is limited, but we can get a sense of the impact of certain predictors by taking the top five covariates with either the greatest importance (for the random forest) or largest coefficients\footnote{Recall that all variables were centered and scaled, making this a valid comparison.} (for the linear models) for each model and tallying how frequently each is included. Offscreen time is included as the first or second most significant variable in the models for all eleven load metrics, a fact foreshadowed by the strong correlation noted in Section \ref{subsec:player_load_metrics}. Offscreen distance and position are both top variables in eight of the eleven models, while the velocity and acceleration entering and leaving the camera window are in the top five for four of the models, primarily the distance metrics. All remaining predictors occur in the top five for just three or fewer of the load metrics.

%considering how frequently they appear as one of the most ``important'' variables across the various models, whether that be according to the importance metric in the case of the random forest or the coefficient size in the case of the linear models . In order to assess the impact of the predictors across the different player load metrics, 

\section{Practical Applications} \label{sec:practical_applications}

Broadcast-derived tracking data has tremendous potential and in this work we have demonstrated its viability for use when evaluating player load in soccer. Because we used out-of-the-box statistical methods with no modification, this type of modeling is widely accessible and should be relatively easy to adopt in practice. Examination of RMSPE and CV values shows that in general, predictions for the various load metrics are very accurate. Some of the stratified metrics, i.e., time in velocity band $[5.7, \infty)$ and very high speed distance, have larger CV values due to the small amount of time players spend in these states, but the RMSPE in these cases is still low enough for use in a practical setting. One consideration when applying our results is how estimate accuracy varies by position and across players. For example, when considering total distance, the RMSPE for defenders, midfielders, and forwards is 216, 164, and 160, respectively. This is driven primarily by differences in censoring rates, with forwards and midfielders being censored just 43.4 and 35.4 percent of the time, versus 52.5 percent of the time for defenders. Censoring rates also vary significantly from player to player, ranging from 19.7 to 60.04 percent of data censored. Accounting for differences in censoring can increase prediction accuracy and improve the efficacy of using broadcast-derived tracking data for external load metric estimation in soccer.

\begin{landscape}
% Table created by stargazer v.5.2.2 by Marek Hlavac, Harvard University. E-mail: hlavac at fas.harvard.edu
% Date and time: Thu, Sep 12, 2019 - 22:27:48
\begin{table}[!htbp] \centering 

  \caption{RMSPE and CV for the base model and subtrack level models on each of the responses} 
  \label{tab:subtrack_results} 
\begin{tabular}{@{\extracolsep{5pt}} |l|rrrrrrrr|} 
% \\[-1.8ex]\hline 
\hline \\[-1.8ex] 
 & \multicolumn{2}{c}{\textbf{Base model}} & \multicolumn{2}{c}{\textbf{Linear model}} & \multicolumn{2}{c}{\textbf{Linear model w/ int}} & \multicolumn{2}{c|}{\textbf{Random forest}} \\
 & RMSPE & CV & RMSPE & CV & RMSPE & CV & RMSPE & CV \\ 
\hline \\[-1.8ex] 
total distance (m) & 288.2 & 0.08 & 202.0 & 0.06 & 188.3 & 0.05 & \textbf{183.0} & \textbf{0.05} \\ 
high speed distance (m) & 164.5 & 0.22 & 113.8 & 0.15 & \textbf{106.0} & \textbf{0.14} & 113.4 & 0.15 \\ 
very high speed distance (m) & 60.4 & 0.44 & 53.4 & 0.39 & 53.3 & 0.39 & \textbf{42.8} & \textbf{0.31} \\ 
time in velocity band $[0, 3.5)$ (s) & 49.9 & 0.03 & 30.4 & 0.02 & \textbf{29.1} & \textbf{0.02} & 29.8 & 0.02 \\ 
time in velocity band $[3.5, 5.7)$ (s) & 37.8 & 0.22 & 26.4 & 0.15 & \textbf{24.5} & \textbf{0.14} & 26.4 & 0.15 \\ 
time in velocity band $[5.7, \infty)$ (s) & 8.8 & 0.43 & 7.9 & 0.38 & 7.9 & 0.38 & \textbf{6.4} & \textbf{0.31} \\ 
total acceleration ($m/s^2$) & 2473 & 0.11 & 1448 & 0.07 & \textbf{1365} & \textbf{0.06} & 1366 & 0.06 \\ 
acceleration density ($m/s^2$) & 0.140 & 0.12 & \textbf{0.113} & \textbf{0.10} & 0.129 & 0.11 & 0.119 & 0.11 \\ 
time in acceleration band $[0.65, 1.46)$ (s) & 34.9 & 0.06 & 25.8 & 0.05 & \textbf{25.7} & \textbf{0.05} & 29.5 & 0.05 \\ 
time in acceleration band $[1.46, 2.77)$ (s) & 45.3 & 0.13 & 30.6 & 0.09 & \textbf{27.8} & \textbf{0.08} & 29.5 & 0.05 \\ 
time in acceleration band $[2.77, \infty)$ (s) & 36.5 & 0.22 & 21.5 & 0.13 & \textbf{20.9} & \textbf{0.13} & 22.3 & 0.13  \\ 
\hline %\\[-1.8ex] 
\end{tabular} 
\end{table} 

% Table created by stargazer v.5.2.2 by Marek Hlavac, Harvard University. E-mail: hlavac at fas.harvard.edu
% Date and time: Thu, Sep 12, 2019 - 22:27:48
\begin{table}[!htbp] \centering 

  \caption{RMSPE and CV for the base model and game level models on each of the responses} 
  \label{tab:game_results} 
\begin{tabular}{@{\extracolsep{5pt}} |l|rrrrrrrr|} 
% \\[-1.8ex]\hline 
\hline \\[-1.8ex] 
 & \multicolumn{2}{c}{\textbf{Base model}} & \multicolumn{2}{c}{\textbf{Linear model}} & \multicolumn{2}{c}{\textbf{Linear model w/ int}} & \multicolumn{2}{c|}{\textbf{Random forest}} \\
 & RMSPE & CV & RMSPE & CV & RMSPE & CV & RMSPE & CV \\ 
\hline \\[-1.8ex] 
total distance (m) & 288.2 & 0.08 & 257.0 & 0.08 & 267.1 & 0.08 & 286.5 & 0.08 \\ 
high speed distance (m) & 164.5 & 0.22 & 150.6 & 0.20 & 153.1 & 0.21 & 159.1 & 0.21 \\ 
very high speed distance (m) & 60.4 & 0.44 & 60.4 & 0.44 & 61.7 & 0.45 & 68.1 & 0.50 \\ 
time in velocity band $[0, 3.5)$ (s) & 49.9 & 0.03 & 41.4 & 0.02 & 42.0 & 0.02 & 60.3 & 0.03 \\ 
time in velocity band $[3.5, 5.7)$ (s) & 37.8 & 0.22 & 34.8 & 0.20 & 35.2 & 0.20 & 37.5 & 0.22 \\ 
time in velocity band $[5.7, \infty)$ (s) & 8.8 & 0.43 & 9.0 & 0.43 & 9.1 & 0.44 & 9.91 & 0.48 \\ 
total acceleration ($m/s^2$) & 2473 & 0.11 & 1748 & 0.08 & 1658 & 0.08 & 2227 & 0.10 \\ 
acceleration density ($m/s^2$) & 0.140 & 0.12 & 0.126 & 0.11 & 0.140 & 0.12 & 0.156 & 0.14 \\ 
time in acceleration band $[0.65, 1.46)$ (s) & 34.9 & 0.06 & 25.7 & 0.05 & 27.4 & 0.05 & 38.1 & 0.07 \\ 
time in acceleration band $[1.46, 2.77)$ (s) & 45.3 & 0.13 & 32.3 & 0.09 & 31.7 & 0.09 & 40.4 & 0.12 \\ 
time in acceleration band $[2.77, \infty)$ (s) & 36.5 & 0.22 & 26.7 & 0.16 & 25.3 & 0.15 & 29.2 & 0.17  \\ 
\hline %\\[-1.8ex] 
\end{tabular} 
\end{table} 

\end{landscape}

\bibliographystyle{apa}
\bibliography{broadcast_tracking_data}

\end{document}